\documentclass[11pt,twoside]{article}  
\usepackage{asp2006}
\usepackage{adassconf}

\begin{document}   

%
%

\paperID{O11.3}

%

\title{Visualization-Directed Interactive Model-Fitting to Spectral Data Cubes}

%
%
%
%
%

\markboth{Fluke, English, Barnes}{Visualization-Directed Interactive Model-Fitting}

%
%
%
%

\author{Christopher J.\ Fluke}
\affil{Centre for Astrophysics \& Supercomputing, Swinburne University 
of Technology, Hawthorn, Victoria, Australia}
\author{Jayanne English}
\affil{Department of Physics and Astronomy, University of Manitoba,
Winnipeg, Manitoba, Canada}
\author{David G.\ Barnes}
\affil{Centre for Astrophysics \& Supercomputing, Swinburne University 
of Technology, Hawthorn, Victoria, Australia}

%

\contact{Christopher Fluke}
\email{cfluke@swin.edu.au}

%
%
%

\paindex{Fluke, C.~J.}
\aindex{English, J.}     
\aindex{Barnes, D.~G.}     

%



\begin{abstract}          
Spectral datasets obtained at radio frequencies and optical/IR wavelengths
are increasing in complexity as new facilities and instruments come online, 
resulting in an increased need to visualize and quantitatively analyze 
the velocity structures. As the visible structure in spectral data cubes 
is not purely spatial, additional insight is required to relate 
structures in 2D space plus line-of-sight velocity to their true
three-dimensional (3D) structures.  This can be achieved through the 
use of models that are converted to velocity-space representations. 
We have used the S2PLOT programming library to enable intuitive, 
interactive comparison between 3D models and spectral data, with 
potential for improved understanding of the spatial configurations. 
We also report on the use of 3D Cartesian shapelets to support 
quantitative analysis. 
\end{abstract}

%
%

\section{Interactive Visualization of Spectral Data Cubes}
Spectral data cubes obtained at radio frequencies (e.g. via aperture synthesis,
mosiacs or multibeam instruments) and optical/infrared wavelengths
(e.g. via integral field units, scanning Fabrey-Perot interferometers, or
multi-object spectrographs) comprise two spatial dimensions and 
one spectral dimension.   In both regimes, the spectral dimension is 
a proxy for the relative line-of-sight velocity between the observer and 
source, resulting in spectral features that are Doppler-shifted from
their rest frequency.  The challenge is to intuitively understand
the relationship between the two spatial plus one velocity observations,
and the true three-dimensional spatial plus three-dimensional velocity
structure.

Solutions exist for simple velocity structures, such as rotating
extragalactic H{\small I} disks, which can be fit by a set of inclined, 
differentially rotating annuli as a function of radius (e.g. 
Rogstad et al. 1974, Begeman 1989).   These models, however,
do not easily account for kinematic features such as warps,
anomalous gas, or mergers, and are not appropriate for interpreting
the complex velocity structures that occur in observations 
of galactic-plane neutral hydrogen. 

\begin{figure}[t]
\epsscale{0.8}
\plotone{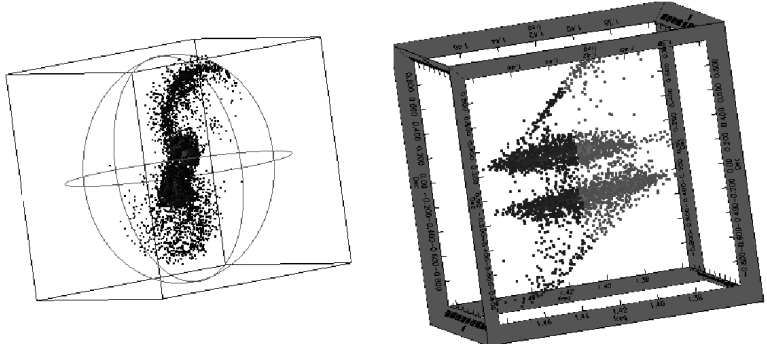}
\caption{ Mapping from simulation space (left) to spectral cube space (right)
using the orientation of a virtual camera to determine the mapping.  Data
is from an $N$-body simulation of a galaxy merger. The spectral cube has
been oriented independently to show velocity-space features. }
\label{O11.3-fig-1}
\end{figure}

One option that presents itself is to generate particle models, 
where the full 3D spatial, ${\mathbf x} = (x,y,z)$, and 3D velocity,
${\mathbf v} = (v_x,v_y,v_z)$, information is known for 
each particle, and project this to spectral cube ``space'', $(x',y',v_z')$,
for display.
The orientation of the model data with respect to the world camera,
with unit up vector, ${\hat \mathbf{u}} = (u_x,u_y,u_z)$, 
view direction, ${\hat \mathbf{w}} = (w_x, w_y, w_z)$, 
and right vector, ${\hat \mathbf{r}} = {\hat \mathbf{w}} 
\times {\hat \mathbf{u}} = (r_x, r_y, r_z)$, is used to obtain the
coordinates for plotting points in spectral cube space:
\begin{eqnarray}
x' & = & x r_x + y r_y + z r_z, \\
y' & = & x u_x + y u_y + z u_z,  \; \mbox{and} \\
v_z' & = & -(v_x w_x + v_y w_y + v_z w_z) + v_{\rm sys},
\end{eqnarray}
where $v_{\rm sys}$ is the overall system velocity. It is straightforward to 
convert $v_z'$ to a frequency or wavelength.  The result of this 
process is demonstrated in Figure \ref{O11.3-fig-1} for a galaxy merger.  

\begin{figure}[t]
\epsscale{0.9}
\plotone{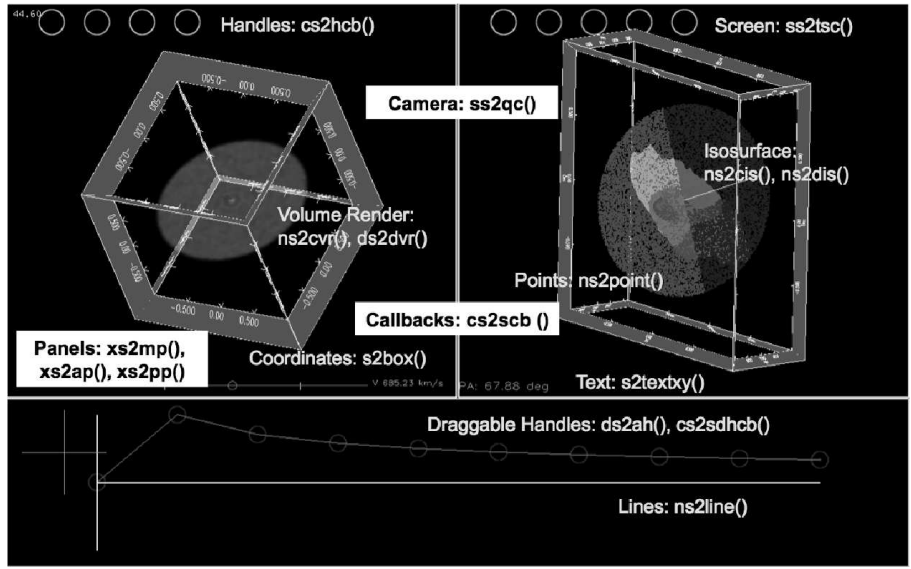}
\caption{Screenshot from a prototype S2PLOT application for interactive
visualization of model and spectral cube data.  The top left panel is
model space, the top right panel is spectral cube space, and the bottom
panel is an interactive rotation curve.  Annotation on the figure gives
the main S2PLOT functionality we are using. The model in this example is a 
rotating disk.}\label{O11.3-fig-2}
\end{figure}

We are using the S2PLOT 3D-graphics library (Barnes et al. 2006) to
develop a framework where models can be examined interactively alongside
their spectral-cube projections.  This approach enables an 
increased intuitive understanding of complex kinematical 
structures.  As a simple, yet powerful application development library
(C/C++/Fortran), S2PLOT provides customizable user interaction via 
keyboard and mouse controls;  output to mono- and stereoscopic displays; 
and support for publication in 3D-PDF format (e.g. Barnes \& Fluke 2008) 
via VRML-export with a single key press.  
Simulation inputs and spectral cube data formats can be handled 
with custom code or standard C libraries (e.g. {\tt cfitiso}
for FITS files).  Figure \ref{O11.3-fig-2} shows a screenshot 
from a prototype application 
-- screen elements are described in the caption, and image annotations
identify the main S2PLOT functionality we use.

We note our S2PLOT approach is limited by the workstation's graphics memory, 
and may not be suitable for very large spectral data cubes ($> 512^3$ voxels). 
To this end, we anticipate integrating our approach with a graphics 
processing unit (GPU) based volume rendering system (see Hassan et al. 2009).  
Alternatively, it may be sufficient to select sub-volumes to analyze.  

We intend to include support for the following basic functionality:
generation of standard model components such as rotating disks, 
warps, hot-spots, expanding shells, $N$-body simulated data;
model-specific controls (e.g. interactive rotation curves and density
profiles for disk models); realistic noise; overlay of simulated spectral
data on observational data; and integrated quantitative 
analysis tools, such as 3D Cartesian shapelet decomposition. 

\section{Quantitative Analysis with 3D Cartesian Shapelets}
\begin{figure}[t]
\epsscale{0.9}
\plotone{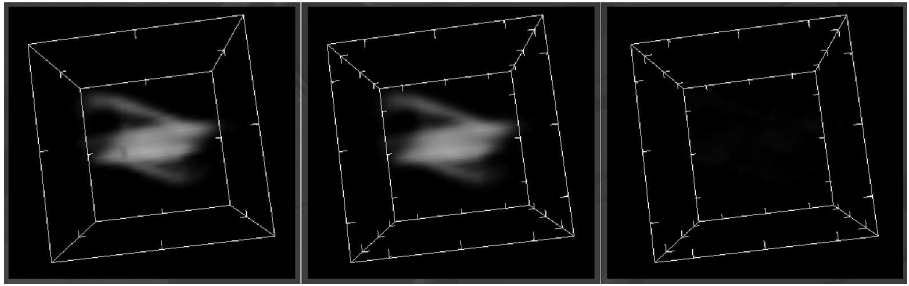}
\caption{Shapelet decomposition of a spectral data cube with $57^3$ voxels.
(Left) input cube; (middle) shapelet reconstruction; (right) residual
cube.  The basic features of the input cube are well-recovered.} \label{O11.3-fig-3}
\end{figure}

The one- and two-dimensional Cartesian shapelets formalism was
introduced by Refregier (2003).  In one dimension, shapelets form an 
orthonormal basis set: 
\begin{eqnarray}
\phi_n(x) & \equiv & (2^n \pi^{1/2} n!)^{-1/2} H_n(x) e^{-x^2/2} \\
B_n(x; \beta) & \equiv & \beta^{-1/2} \phi_n(\beta^{-1} x)
\end{eqnarray}
where $n$ is the shapelet order, $H_n(x)$, is the $n$-th Hermite polynomial,
and $\beta$ is a scale factor.  We have recently extended the Cartesian 
approach to 3D, deriving a number of important analytic relationships 
(Fluke et al. in prep). While a full presentation is beyond the scope of this 
paper, the orthonormality of shapelet states enables us to write 3D shapelets 
in terms of functions:
\begin{equation}
B_{3,{\mathbf n}}({\mathbf x};\beta) \equiv \beta^{-3/2} 
\phi_{n_1} (\beta^{-1}x) \phi_{n_2} (\beta^{-1}y) \phi_{n_3} (\beta^{-1}z), 
\end{equation}
where ${\mathbf n} = (n_1, n_2, n_3)$ are the integer shapelet orders.
An arbitrary (sufficiently well-behaved) 3D structure, $f_3({\mathbf x})$, 
can be decomposed into a weighted sum of shapelet coeffecients, 
$f_{3,{\mathbf n}}$, by integration over a volume, $V$:
\begin{equation}
f_{3,\mathbf{n}} = \int_V f_3(\mathbf{x}) B_{3,\mathbf{n}} (\mathbf{x};\beta) {\rm d}^3 x
\end{equation}
subject to $n_1 + n_2 + n_3 \leq n_{\rm max}$, and an ``optimal'' $\beta$-value. The shapelet reconstruction is:
\begin{equation}
\hat{f}_3(\mathbf{x}) = \sum_{n_1,n_2,n_3}^{n_{\rm max}} \alpha(\mathbf{n}) f_{3,\mathbf{n}} B_{3,\mathbf{n}}
(\mathbf{x}; \beta),
\end{equation}
where $\alpha(\mathbf{n})$ is an optional filter term.
An example of the decomposition and reconstruction process for a mock spectral
data cube (no noise) is shown in Figure \ref{O11.3-fig-3}.  As a highly parallel task,
the decomposition and reconstruction processes can be efficiently 
implemented on a GPU (Barsdell et al. 2009).  

\section{Conclusion}
We are using the S2PLOT programming library to enable interactive, visualization-directed 
model-fitting of spectral data cubes, leading to improved intuitive understanding 
of complex kinematic structures.  Shapelet space provides new opportunities for 
quantitative analysis of complex 3D structures.  We intend a public code release on completion.

\end{document}